%% file: main.tex
\newif\ifdraft
\drafttrue

\documentclass[sigconf,authorversion,nonacm]{acmart}

\usepackage{amsmath,amsfonts}
\usepackage{algorithmic}
\usepackage{graphicx}
\usepackage{textcomp}
\usepackage{xcolor}
\usepackage{listings}
\usepackage{textcomp}
\usepackage{siunitx}
\usepackage{multirow}
\usepackage{xspace}
\usepackage{balance}
\usepackage{caption,subcaption}



\usepackage[T1]{fontenc}

\definecolor{backgroundColour}{HTML}{F8F8F8}
\definecolor{keywordclr}{HTML}{3969AC}
\definecolor{commentclr}{HTML}{A2A0AA}
\definecolor{stringsclr}{HTML}{11A579}
\definecolor{fnctionclr}{rgb}{0.467, 0, 0.533}
\definecolor{builtinclr}{rgb}{0.35, 0, 0.533}
\definecolor{symbolsclr}{rgb}{0.5, 0.25, 0.25}   
\definecolor{numbersclr}{rgb}{0.8, 0.2, 0}
\definecolor{bckgrndclr}{rgb}{0.91, 0.95, 0.95}
\lstdefinestyle{PythonStyle}{
    language=Python,
    backgroundcolor=\color{backgroundColour},
    keywordstyle=\color{keywordclr}\bfseries,
    stringstyle=\color{stringsclr},
    commentstyle=\color{commentclr}\itshape,
    upquote=true,
    basicstyle=\ttfamily\linespread{0.9}\footnotesize,
    breakatwhitespace=false,
    breaklines=true,
    captionpos=b,
    keepspaces=true,
    numbers=left,
    numbersep=5pt,
    numberstyle=\color{commentclr}\ttfamily\tiny,
    showspaces=false,
    showstringspaces=false,
    showtabs=false,
    tabsize=2,
    xleftmargin=1.25em,
    frame=single,
    framexleftmargin=1.25em,
    morekeywords={assert,with,as}
}
\ifdraft
  \newcommand{\todo}[1]{\textcolor{red}{ TODO: #1 }}
  \newcommand{\note}[1]{\textcolor{blue}{ Note: #1 }}
  \newcommand{\ian}[1]{\textcolor{green}{ Ian: #1 }}
  \newcommand{\greg}[1]{\textcolor{purple}{ Greg: #1 }}
  \newcommand{\val}[1]{\textcolor{orange}{ Valerie: #1 }}
  \newcommand{\kyle}[1]{\textcolor{teal}{ Kyle: #1 }}
\else
  \newcommand{\todo}[1]{}
  \newcommand{\note}[1]{}
  \newcommand{\ian}[1]{}
  \newcommand{\greg}[1]{}
  \newcommand{\val}[1]{}
  \newcommand{\kyle}[1]{}
\fi


\begin{document}

\title{Accelerating Python Applications with Dask and ProxyStore}

\author{J. Gregory Pauloski}
\affiliation{%
  \institution{University of Chicago}
  \institution{Argonne National Laboratory}
  \city{Chicago}
  \state{Illinois}
  \country{USA}
}

\author{Klaudiusz Rydzy}
\affiliation{%
  \institution{Loyola University Chicago}
  \city{Chicago}
  \state{Illinois}
  \country{USA}
}

\author{Valerie Hayot-Sasson}
\affiliation{%
  \institution{University of Chicago}
  \institution{Argonne National Laboratory}
  \city{Chicago}
  \state{Illinois}
  \country{USA}
}

\author{Ian Foster}
\affiliation{%
  \institution{University of Chicago}
  \institution{Argonne National Laboratory}
  \city{Chicago}
  \state{Illinois}
  \country{USA}
}

\author{Kyle Chard}
\affiliation{%
  \institution{University of Chicago}
  \institution{Argonne National Laboratory}
  \city{Chicago}
  \state{Illinois}
  \country{USA}
}

\renewcommand{\shortauthors}{Pauloski, Rydzy, Hayot-Sasson, Foster, and Chard}

\begin{abstract}
Applications are increasingly written as dynamic workflows underpinned by an execution framework that manages asynchronous computations across distributed hardware.
However, execution frameworks typically offer one-size-fits-all solutions for data flow management, which can restrict performance and scalability.
ProxyStore, a middleware layer that optimizes data flow via an advanced pass-by-reference paradigm, has shown to be an effective mechanism for addressing these limitations.
Here, we investigate integrating ProxyStore with Dask Distributed, one of the most popular libraries for distributed computing in Python, with the goal of supporting scalable and portable scientific workflows.
Dask provides an easy-to-use and flexible framework, but is less optimized for scaling certain data-intensive workflows.
We investigate these limitations and detail the technical contributions necessary to develop a robust solution for distributed applications and demonstrate improved performance on synthetic benchmarks and real applications.
\end{abstract}


\maketitle

\section{Introduction}
\label{sec:introduction}

Contemporary computational science applications executed at scale are increasingly written as workflows, collections of many distinct tasks interconnected through data dependencies.
This trend has necessitated the development of advanced computational tools and frameworks that can glue software components together and provide a platform for scalable and flexible execution on arbitrary hardware.
Dynamic workflow execution frameworks, including Dask~\cite{rocklin2015dask}, Dragon~\cite{radcliffe2023dragon}, Parsl~\cite{babuji19parsl}, Ray~\cite{moritz2018ray}, and TaskVine~\cite{slydelgado2023taskvine}, have emerged as powerful solutions to this challenge in the high-performance Python community.
Applications can be expressed as fine-grained tasks, typically a function, with special constructs, such as futures, used to implicitly express inter-task data dependencies.
The framework then abstracts the complexities of executing tasks in parallel and managing intermediate data across personal, cloud, or high-performance computing (HPC) systems.

However, this class of systems, which typically use one-size-fits-all solutions to facilitate intermediate data movement, often fail to meet the data flow needs of modern, dynamic, and data-intensive applications.
Workflow systems commonly rely on a shared file system, as in Parsl and TaskVine, or peer-to-peer TCP communication, as in Dask Distributed, due to simplicity and availability of these approaches.
Thus, workflow systems, and therefore applications, often fail to take advantage of advanced technologies available or suffer from functional but sub-optimal solutions.

Recent work has used the transparent object proxy paradigm to decouple data flow complexities from control flow-optimized execution frameworks~\cite{pauloski2023proxystore,pauloski2024proxystore}.
In this case, proxies function as lightweight, wide area references to objects located in arbitrary data stores.
Proxies can be communicated cheaply and are resolved just-in-time via performant bulk transfer methods in a manner which is transparent to the consumer code.
ProxyStore, a Python framework that implements this paradigm, has found success in many scientific domains and with many distributed execution frameworks~\cite{ward2021colmena,harb2023uncovering,collier2023developing,dharuman2023protein,ward2023colmena}.

Here, we investigate how to build scalable and portable scientific workflows through the careful integration of ProxyStore with Dask Distributed.
Dask, with 12.3k stars on GitHub\footnote{Dask GitHub: \url{https://github.com/dask/dask}.} and 4.9M downloads per month in September 2024,\footnote{Dask Distributed PyPI downloads: \url{https://pypistats.org/packages/distributed}.} is one of the most popular libraries for distributed and parallel computing in Python.
We first give a brief overview of Dask Distributed, ProxyStore, and related efforts in optimizing data flow.
We then investigate limitations of Dask's data management model and detail the technical contributions necessary to overcome these limitations with ProxyStore.
The result is a robust and easy-to-use solution for building sophisticated computational science workflows, which we demonstrate through synthetic performance evaluations
and real-world applications.

\section{Background and Motivation}
\label{sec:background}

\textbf{Dask}~\cite{rocklin2015dask} is a parallel computing library in Python that enables efficient parallel computations on large datasets by breaking them down into smaller, manageable tasks.
Dask Distributed extends the Dask and Python \texttt{concurrent.futures} APIs to provide a lightweight and easy-to-use library for distributed computing.
A centralized scheduler manages the dynamic execution of tasks across local cores or multiple nodes in a cluster and is optimized for low-latency task dispatching, spending between a 100~$\mu$s and 1~ms on each task.
However, this overhead can drastically increase when the graph of a task is large, such as when task parameters are large or complex.
Large task graphs can incur significant I/O overheads in the scheduler for serialization, communication, and deserialization of messages.


Dask provides mechanisms to optimize data transfer:
(1) array-like data can be scattered and gathered directly across workers;
(2) native interfaces optimize common data operations
\footnote{\url{https://docs.dask.org/en/stable/best-practices.html\#load-data-with-dask}}
(e.g., through Dask Arrays, Bags, DataFrames, and Delayed);
and
(3) objects already located on workers, such as the results of tasks, will be communicated directly between workers rather than through the scheduler.
The goal of these solutions is to prefer passing task data by reference rather than embedding data directly in the graph; however, these solutions do not cover all data types or application structures.
For example, frequently moving large objects between the client and workers is considered an anti-pattern; Dask prefers that data remain on the worker cluster.
Yet, this is a common pattern in scientific applications (e.g., active learning~\cite{ward2021colmena,ward2023colmena}) that is not supported as well by Dask.

\input{figures/schematic}

\textbf{ProxyStore}~\cite{pauloski2023proxystore} is a library that facilitates efficient data flow management in distributed Python applications.
The transparent object proxy, a reference-like object, is the core building block of ProxyStore, and, unlike traditional references that are only valid within the virtual address space of a single process, the proxy refers to an object in distributed storage and can be implicitly dereferenced in arbitrary processes, even on remote machines.
The proxy is transparent in that it dereferences its target object when used---referred to a just-in-time resolution---and it forwards all operations on itself to its target object.
This paradigm results in the best of both pass-by-reference and pass-by-value semantics.

A proxy is initialized with a factory, a self-contained callable object that is invoked when the proxy is resolved to retrieve the target object.
This self-contained nature and transparency of the proxy means a consumer is not aware of the low-level communication mechanisms used by the proxy; rather, this is unilaterally determined by the producer of the proxy.
This paradigm improves performance and portability by reducing transfer overheads through intermediaries, abstracting low-level communication methods, and reducing code-complexity.

ProxyStore separates the high-level interface, the \texttt{Store}, responsible for creating proxies from the low-level interface, the \texttt{Connector}, responsible for interfacing with the byte-level mediated communication and storage channels.
Connectors to many storage systems (object stores, shared file-systems) and transfer protocols (Grid FTP, TCP, RDMA, and WebRTC) are provided.
These interfaces have been used to build high-level patterns for distributed futures, object streaming, and distributed memory management~\cite{pauloski2024proxystore}.

\textbf{Related work} has investigated ways to improve the performance of data-intensive workflows with Dask.
Dask provides a UCX communication protocol implementation as an alternative to the TCP default to leverage advanced networking technologies such as Inifiniband or NVLink.
Later work developed an MPI-based communication interface for Dask for GPU-accelerated programs~\cite{shafi2021efficient}.
TaskVine~\cite{slydelgado2023taskvine}, a distributed workflow engine that exploits node-local storage to optimize task placement and execution, and  Ray~\cite{moritz2018ray}, a popular distributed computing library, provide alternative scheduler implementations for Dask workflows.
These solutions can yield considerable performance gains in certain applications but also have limited deployment scenarios.
ProxyStore, in contrast, provides more fine-grained data flow customization with wider support for communication protocols and storage mediums.

\section{Integration Model}
\label{sec:integration}

ProxyStore can alleviate data transfer overheads in Dask by proxying large task objects instead of embedding them directly in the task graph (\autoref{fig:schematic}).
Importantly, use of ProxyStore does not require modification to task code and is not mutually exclusive with Dask optimization options.
Here, we discuss the methods for integrating ProxyStore into Dask applications, the technical challenges overcome to ensure compatibility and performance, and the features added to support development of robust applications.

\input{listings/code}

\textbf{Methods:}
There are three methods, exemplified in \autoref{lst:code-examples}, to integrate ProxyStore into a Dask application: (1) manually proxy objects using ProxyStore's existing tooling, (2) use our custom Dask client to automatically proxy objects; or (3) use our custom executor interface to intelligently proxy objects and manage memory.
For simple applications, the manual approach works well, but it can require significant code changes in more sophisticated applications.
The custom client provides a drop-in replacement for existing Dask applications.
The \texttt{StoreExecutor}, which extends Python's \texttt{concurrent.futures} interface, is the most powerful approach: it is compatible with many other executor client types, such as those provided by Parsl and TaskVine; custom policies can be defined to determine what objects to automatically proxy and, when combined with ProxyStore's \texttt{MultiConnector}, what mediated storage option to use; and it uses ProxyStore's ownership model~\cite{pauloski2024proxystore}, inspired by Rust's ownership and borrowing semantics, to perform safe and automatic memory management of proxies. 

\textbf{Compatibility:}
Dask performs introspection on objects included in task graphs to enable optimizations.
Each task has an associated key, a hash of the function and arguments.
The scheduler uses the key to reuse results of previously computed pure functions (functions that always return the same result given the same inputs).
Similarly, Dask inspects task object types to apply specialized serializers.
These optimizations enhance performance but interacted poorly with proxy types.
For example, Dask serialization would crash when accessing the \texttt{\_\_module\_\_} property of a proxy, and hashing or checking the type of a proxy would resolve the proxy, incurring unexpected I/O costs.
We resolved this by creating a custom implementation of Python's \texttt{@property} decorator and modified the proxy to cache common read-only attributes of a proxied object.
These include the module path, the class type, and the hash of the target object to ensure that a proxy need not be resolved when Dask accesses common object metadata.

\textbf{Performance:}
We improved ProxyStore's performance for scientific workloads by overhauling the serialization system to minimize memory copies and support custom serialization mechanisms for specific data types.
We have provided initial support for NumPy arrays, Pandas DataFrames, and Polars DataFrames.
Serialization of these types is 2--3$\times$ faster compared to pickle, which ProxyStore previously used.

We also extend ProxyStore to support Distributed Asynchronous Object Storage (DAOS) as a mediated storage system.
DAOS is a distributed object store designed for high-speed non-volatile memory like Intel Optane~\cite{liang2020daos} and NVMe and is available on next-generation compute clusters like Aurora at the Argonne Leadership Computing Facility.
DAOS is typically deployed across a machine in a similar fashion to a shared file system like Lustre.
Thus, using the DAOS within ProxyStore is easy---minimal configuration is required---and performance is superior to shared file systems.
The user need only provide the name of their DAOS pool and container to use.

\textbf{Robustness:}
The introduction of type hints and static type checkers such as mypy has significantly improved code quality and maintainability, ultimately leading to more robust software.
The proxy model, however, relies strongly on Python's duck typing and, thus, code that uses ProxyStore cannot be statically analyzed and validated, leading to often cryptic errors when a proxy type is used incorrectly at runtime.
We created a mypy extension that can statically infer usage of proxy types.
For example, mypy can understand that any attributes or methods on a type \texttt{T} are also available on a \texttt{Proxy[T]} or that a function that accepts a \texttt{ProxyOr[T]} should work with a  \texttt{T} or \texttt{Proxy[T]}.
This tool ensures that scientific software developers can write code that will work with and without ProxyStore, improving code compatibility and maintainability.

\section{Evaluation}

We used the TaPS benchmark suite~\cite{pauloski2024taps} to evaluate the performance benefits of using ProxyStore within Dask applications.
Experiments were performed using the Sunspot system at the Argonne Leadership Computing Facility.
\footnote{This work was done on a pre-production supercomputer with early versions of the Aurora software development kit.}
Sunspot has \num{128} nodes interconnected by an HPE Slingshot 11 network and a high-performance DAOS storage system.
Each node contains two Intel Xeon Max CPUs with 52 physical cores, 64 GB of high-bandwidth memory, 128 GB of DDR5 memory per CPU, and six Intel Data Center Max GPUs.
We used Python 3.11, Dask Distributed 2024.7.1, ProxyStore 0.7.1, ProxyStore Extensions v0.1.4, and TaPS 0.2.1.
Analysis, code, and results are available at \url{https://github.com/proxystore/hppss24-demo}.

We performed ProxyStore experiments using Redis due to DAOS outages on Sunspot at the time of writing; a Redis server was started on the rank 0 node of each batch job.
Configuring ProxyStore to use DAOS would be even easier, as described in \autoref{sec:integration}, and we expect comparable performance outcomes due to DAOS leveraging NVMe storage distributed throughout the racks of the cluster.

\input{figures/overheads}
\textbf{Overheads:}
We measure the round-trip time of no-op tasks with payloads of varying sizes in \autoref{fig:overheads}.
This experiment represents a worst-case scenario for the Dask scheduler: all data is sent between the client and workers and no data is reused across multiple tasks.
Using ProxyStore's pass-by-proxy model improves round-trip time for larger task payloads ($>100$~kB) by up to 50\%.
This improvement is attributed to (1) smaller messages to be serialized and communicated, (2) less data transferred through the scheduler, and (3) improvements to ProxyStore's serialization that reduce memory copies.

\input{figures/scaling}

\textbf{Scaling:}
We measure task throughput with and without ProxyStore as a function of the number of Dask workers $n$.
Each node hosts up to 104 workers, the number of physical cores per node.
We execute \num{10000} tasks that consume and produce 1~MB of random data (chosen based on the results in \autoref{fig:overheads}).
An initial batch of $n$ tasks are submitted; as current tasks complete, new tasks are submitted until all tasks are finished.
Tasks are essentially no-ops besides the result data generation which takes only $O(1)$~ms; thus, the goal of this experiment is to stress the Dask scheduler and understand its limits.
As depicted in \autoref{fig:scaling-performance}, task throughput with Dask quickly plateaus around 170 tasks per second and degrades when utilizing 104 workers.
Use of ProxyStore alleviates data transfer burdens from the scheduler, enabling higher sustained throughput; however we still observe the same drop in performance at 104 workers which may indicate the presence of performance limitations in the Dask scheduler that are independent of data volume.

\input{figures/apps}

\textbf{Applications:}
We use three reference applications from TaPS representing an array of data patterns.
Cholesky decomposition has short tasks that consume and produce large arrays, federated learning has long tasks that consume and produce large models, and molecular design has short tasks that consume and produce small datasets and models.
We chose these three applications because they are implemented in a manner which accentuates data transfer between the client and workers.
As demonstrated in \autoref{fig:app-performance}, ProxyStore yields the greatest benefits to Dask applications with larger tasks payloads and shorter running tasks---applications where task overheads represent a larger proportion of overall runtime.

\section{Conclusion}
\label{sec:conclusion}

The pass-by-proxy model of ProxyStore is a viable alternative to data flow management in distributed Dask applications.
We discussed the ways in which Dask applications can be extended with ProxyStore, the diverse storage systems and communications channels supported by ProxyStore, and the technical contributions necessary to make the integration possible.
Experiments show that ProxyStore reduces task overheads when task graphs are large, improves task throughput at scale, and accelerates applications which suffer from I/O bottlenecks in the Dask scheduler.

\section*{Acknowledgment}
This research used resources of the Argonne Leadership Computing Facility, a U.S. Department of Energy (DOE) Office of Science user facility at Argonne National Laboratory and is based on research supported by the U.S. DOE Office of Science-Advanced Scientific Computing Research Program, under Contract No. DE-AC02-06CH11357.

\balance
\bibliographystyle{ACM-Reference-Format}
\bibliography{refs}

\end{document}

%% file: figures/schematic.tex
\begin{figure}
    \centering
    \includegraphics[width=\columnwidth,trim={0 2px 0 2px},clip]{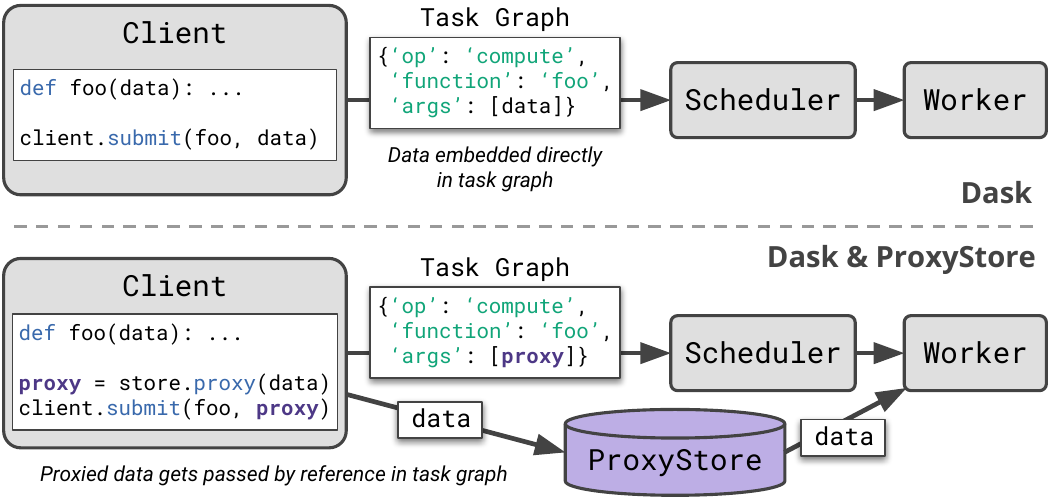}
    \caption{Pass-by-proxy semantics reduce data flow through the Dask scheduler without altering application behavior.}
    \label{fig:schematic}
    \Description[Overview of data flow with Dask and ProxyStore.]{Pass-by-proxy semantics reduce data flow through the Dask scheduler without altering application behavior.}
\end{figure}

%% file: listings/code.tex
\begin{figure}[t]
\begin{subfigure}{\columnwidth}
\begin{lstlisting}[style=PythonStyle]
from dask.distributed import Client
from proxystore.ex.connectors.daos import DAOSConnector
from proxystore.store import Store

client = Client()
connector = DAOSConnector(pool=..., container=...)

with Store('example', connector) as store:
    proxy = store.proxy([1, 2, 3])
    future = client.submit(sum, proxy)
    assert future.result() == 6
\end{lstlisting}
\caption{A proxy can be manually created via the \texttt{Store} interface and passed directly to tasks in place of the actual object.}
\end{subfigure}

\bigskip

\begin{subfigure}{\columnwidth}
\begin{lstlisting}[style=PythonStyle]
from proxystore.ex.plugins.distributed import Client
from proxystore.ex.connectors.daos import DAOSConnector
from proxystore.store import Store

connector = DAOSConnector(pool=..., container=...)

with Store('example', connector) as store:
    client = Client(ps_store=store, ps_threshold=1000)
    future = client.submit(sum, [1, 2, 3])
    assert future.result() == 6
\end{lstlisting}
\caption{The custom Dask Distributed \texttt{Client} will automatically proxy task input and output objects larger than a user-defined threshold (e.g., 1~kB).}
\end{subfigure}

\bigskip

\begin{subfigure}{\columnwidth}
\begin{lstlisting}[style=PythonStyle]
import sys
from dask.distributed import Client
from proxystore.ex.connectors.daos import DAOSConnector
from proxystore.store import Store
from proxystore.store.executor import StoreExecutor

client = Client()
connector = DAOSConnector(pool=..., container=...)

with StoreExecutor(
    client,
    store=Store('example', connector),
    should_proxy=lambda x: sys.getsizeof(x) >= 1000,
) as executor:
    future = executor.submit(sum, [1, 2, 3])
    assert future.result() == 6
\end{lstlisting}
\caption{The \texttt{StoreExecutor} can combine a \texttt{Store} and Dask \texttt{Client} and supports custom policies for what objects should be automatically proxied (here, objects larger than 1~kB) and automatically manages proxy lifetimes.
}
\end{subfigure}
\caption{ProxyStore is easily compatible with existing applications. Here we demonstrate the three integration patterns. The \texttt{DAOSConnector}, introduced in \autoref{sec:integration}, is used, but this specific connector can be exchanged depending on the application requirements and execution environment.
\label{lst:code-examples}
\Description[Code snippets of Dask and ProxyStore usage.]{ProxyStore is easily compatible with existing applications. Here we demonstrate the three integration patterns.}
}
\end{figure}

%% file: figures/overheads.tex
\begin{figure}
    \centering
    \includegraphics[width=\columnwidth,trim={0 0px 0 0px},clip]{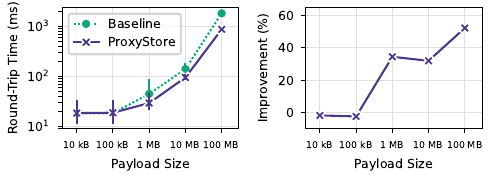}
    \caption{(Left) No-op task round-trip time with various payload sizes. (Right) Relative improvement in round-trip time compared to the baseline when using ProxyStore.}
    \label{fig:overheads}
    \Description[No-op task round-trip time performance.]{(Left) No-op task round-trip time with various payload sizes. (Right) Relative improvement in round-trip time compared to the baseline when using ProxyStore.}
\end{figure}

%% file: figures/scaling.tex
\begin{figure}
    \centering
    \includegraphics[width=\columnwidth,trim={0 0px 0 0px},clip]{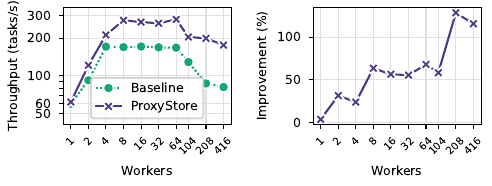}
    \caption{(Left) No-op task throughput with various worker counts. Tasks consume and produce 1~MB of random data. (Right) Relative improvement in throughput compared to the baseline when using ProxyStore. ProxyStore alleviates data flow burdens from the Dask scheduler, enabling the scheduler to dispatch tasks faster.}
    \label{fig:scaling-performance}
    \Description[Scaling performance results.]{(Left) No-op task throughput with various worker counts. Tasks consume and produce 1~MB of random data. (Right) Relative improvement in throughput compared to the baseline when using ProxyStore. ProxyStore alleviates data flow burdens from the Dask scheduler, enabling the scheduler to dispatch tasks faster.}
\end{figure}

%% file: figures/apps.tex
\begin{figure}
    \centering
    \includegraphics[width=\columnwidth,trim={0 0px 0 0px},clip]{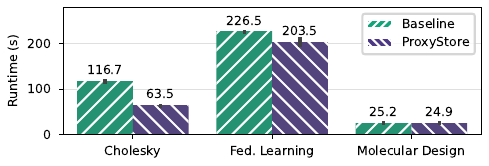}
    \caption{ProxyStore can reduce Dask overheads applications that embed large objects in the task graph, such as the Cholesky decomposition example and federated learning simulation provided by TaPS.}
    \label{fig:app-performance}
    \Description[Application performance results.]{ProxyStore can reduce Dask overheads applications that embed large objects in the task graph, such as the Cholesky decomposition example and federated learning simulation provided by TaPS.}
\end{figure}